\documentclass[12pt]{article}
\usepackage{a4}
\usepackage{amsmath}
\usepackage{amssymb}
\usepackage{hyperref}
\usepackage{mathrsfs}
\usepackage{textcomp}
\usepackage{cite}
\usepackage{bm}
\usepackage[numbers,sort&compress]{natbib}
\newcommand{\be}{\begin{equation}}
\newcommand{\ee}{\end{equation}}
\newcommand{\bea}{\begin{eqnarray}}
\newcommand{\eea}{\end{eqnarray}}

\begin{document}
	\thispagestyle{empty}
	\def\thefootnote{\fnsymbol{footnote}}
	\begin{center}\Large
		A note on Zamolodchikov's recursion relation for the torus conformal block and its light limit \\
		\vskip 2em
		July 2026
	\end{center}\vskip 0.2cm
	\begin{center}
Armen Poghosyan		 $^{1}$\footnote{armenpoghos@yerphi.am}\,,\,
		Hasmik Poghosyan$^{1}$ \footnote{hasmikpoghos@gmail.com}
	\end{center}
	\vskip 0.2cm
	
	\begin{center}
		$^1$ Yerevan Physics Institute, \\
		Alikhanian Br. 2, 0036\, Yerevan\\
		Armenia
	\end{center}

	\vskip 1.5em
	\begin{abstract} \noindent
	In this paper, we review the Zamolodchikov-like recursion relations for torus one-point conformal blocks in both Liouville and $A_2$ Toda theories. 
	Starting from these relations, we derive the corresponding recursion relations in the light asymptotic limit. 
	In Liouville theory, the light-limit recursion reproduces the known expression for the one-point light conformal block in terms of the Gauss hypergeometric function. 
	For $A_2$ Toda theory, our recursion relation provides a new, efficient method for explicit calculations, and we have verified that it is in full agreement with previously known results in the literature.
	
	Last but not least, we present closed-form expressions for the coefficients of the insertion-point expansion for both the generic and light $A_2$ Toda one-point blocks.
\end{abstract}
\vspace{1cm}


\newpage
\tableofcontents
\newpage
\section{Introduction}
In any quantum field theory, the derivation of correlation functions is of primary importance. 
In two-dimensional conformal field theories \cite{BELAVIN1984333}, conformal blocks constitute the fundamental holomorphic building blocks of these correlation functions. 
The study of the quasiclassical limit  \cite{Braaten:1983np,Braaten:1982yn,Thorn:2002am} provides a natural first step toward a complete quantum formulation, and conformal blocks in this regime have therefore been extensively investigated over the past several decades. 
For studies of quasiclassical conformal blocks in Liouville theory and its several generalizations, see 
\cite{Fateev:2007ab,Fateev:2010za,Menotti:2006gc,Poghosyan:2015oua,Poghosyan:2017qdv,Apresyan:2017few,Belavin:2018hfm,Poghosyan:2020zzg,Cipriani:2025ikx,Bykov:2025lhl}.

The main goal of this paper is to study the light limit of the one-point torus block in quantum Liouville \cite{Polyakov:1981xx} and $A_2$ Toda \cite{Zamolodchikov:1985wn} theories. Before turning to this problem, let us briefly review the current state of the art.

We start with the light limit of the four-point block. For Liouville theory its expression is well known, see \cite{Zamolodchikov:1984eqp,Zamolodchikov:1987avt,Mironov:2009qn}; for the light correlation function, see \cite{Hama:2013ama}. The ${\cal N}=1$ super Liouville light blocks were studied in \cite{Poghosyan:2017qdv}, the $A_2$ Toda light block in \cite{Fateev:2011qa}, and its $A_{n-1}$ Toda generalization in \cite{Poghosyan:2016lya}.

Let us now turn to what is known about the one-point light block. In \cite{Hadasz:2009db,Alkalaev:2016fok}, an elegant expression for the Liouville one-point light conformal block was obtained in terms of the hypergeometric function $_2F_1$. 
For $A_2$ Toda theory, an explicit expression was derived in \cite{Belavin:2024nnw} using the shadow formalism. 
Their result is written as a complicated infinite sum involving the hypergeometric function $_3F_2$ evaluated at unit argument, leading to potential subtleties related to analytic continuation. 
Later, using the AGT correspondence \cite{Alday:2009aq,Alba:2010qc,Wyllard:2009hg,He:2012bi,Poghossian:2017atl} and instanton counting 
\cite{Losev:1997bz,Nekrasov:2002qd,Nekrasov:2003rj,Flume:2002az,Bruzzo:2002xf}, an explicit expression for $A_{n-1}$ Toda theory  was found in \cite{Poghosyan:2026yow}. 
This result is valid for arbitrary positive integers $n\ge 2$. 
However, the expression has a drawback: it contains spurious poles.

It is well known that one of the most direct ways to derive a conformal block is through the Zamolodchikov recursion relation \cite{Zamolodchikov:1984eqp,Zamolodchikov:1987avt}. 
Such recursion relations are known for the one-point torus conformal blocks in both Liouville theory and $A_2$ Toda theory 
\cite{Poghossian:2009mk,Poghossian:2017atl}. 
In this paper, using these Zamolodchikov-like recursion relations, we derive analogous  relations for the corresponding light blocks. 
In the Liouville case, from our result one can readily  reproduce the expression of \cite{Hadasz:2009db,Alkalaev:2016fok}. 
For $A_2$ Toda theory, we have checked that our recursion relation agrees with the result of \cite{Poghosyan:2026yow} up to thirtieth order in $q$, defined in (\ref{q}). 
From a practical point of view, our recursion relation is particularly convenient: it contains neither the spurious poles present in \cite{Poghosyan:2026yow} nor the analytic-continuation subtleties associated with the expression of \cite{Belavin:2024nnw}. 
Furthermore, our computations provide an independent check of the general $A_2$ recursion relation of \cite{Poghossian:2017atl}, which is relatively recent and has not yet been tested as extensively.

A nice result closely related to Zamolodchikov's recursion relation was obtained in \cite{Perlmutter:2015iya}, where, from  the recursion, an explicit closed-form expression for an arbitrary expansion coefficient of the generic four-point and torus one-point conformal blocks was derived. In this paper we obtain the analogous expression for the $A_2$ Toda one-point block and also write down its light-limit form.

In the light limit, another expression for an arbitrary expansion coefficient was obtained in \cite{Poghosyan:2026yow}, and it would be interesting to understand how these two expressions are related.

A Zamolodchikov-like recursion relation is also known for the $A_2$ Toda four-point conformal block \cite{Poghossian:2017atl}. A straightforward consequence of our result is that a closed-form expression for its expansion coefficients can be written down in the same way.

Finally, alongside the paper we also provide  a Mathematica notebook containing the following material. 
For Liouville theory, it includes the general Zamolodchikov-like recursion relation for the one-point torus conformal block, the corresponding recursion relation in the light limit, and a check of their consistency. 
For Toda theory, the analogous code is provided as well, together with a check that our result agrees with the expression of \cite{Poghosyan:2026yow}.

The paper is organized as follows. 

In Section~\ref{Lt}, we first review the Zamolodchikov-like $\Delta$-recursion relation for the torus one-point conformal block in Liouville theory. 
Then in Subsection \ref{Llight} we take its light asymptotic limit and derive the corresponding recursion relation for the Liouville light block. 
From this recursion, we recover the known expression for the one-point light conformal block in terms of the Gauss hypergeometric function $_2F_1$. 

In Section~\ref{TodaLight}, we turn to $A_2$ Toda theory. 
We review the relevant notation and the Zamolodchikov-like recursion relation for the generic block, and present our result for the explicit form of its expansion coefficients. 
Then in Subsection \ref{Tlight} we derive the light-limit form of both the recursion and the explicit representation. 
Subsection \ref{WrecLight} contains the main result of this paper, where we write down the final form of the $A_2$ Toda light one-point block recursion relation.

\section{Liouville Theory}
\label{Lt}
\subsection{Zamolodchikov-like $\Delta$-recursion for one-point torus block  in Liouville theory}
\label{LzamRec}

In this section, we consider Liouville theory \cite{Polyakov:1981xx}, a two-dimensional conformal
field theory with Virasoro central charge
\bea
c=1+6Q^2\,,
\qquad
Q=b+b^{-1}\,,
\eea
where $b$ is the dimensionless Liouville coupling. 
We denote the torus one-point conformal block by 
\bea
\label{1pBlockLi}
{\cal F}_{\alpha}^\lambda (q)=q^{\frac{c}{24}-\Delta_{\alpha}} {\rm tr}_{\alpha}
\left(q^{L_0-\frac{c}{24}} V_{\lambda }(1)\right)\,,
\eea
where $\alpha$ and $\lambda$ are the internal and external charge parameters, respectively. 
The conformal dimension of a primary field, say $V_\alpha(x)$, is expressed in terms of its charge parameter as
\bea
\label{dim}
\Delta_{\alpha}=\alpha(Q-\alpha)\,.
\eea
We also use the standard notation
\bea
\label{q}
q=e^{2\pi i\tau}\,,
\eea
where   $\tau$ is the torus  modular parameter.
The torus one-point conformal block can be computed recursively using the
Zamolodchikov-type $\Delta$-recursion relation derived in
\cite{Poghossian:2009mk}; see \cite{Hadasz:2009db} for a more CFT-oriented
account of this recursion. It is convenient to introduce a function related
to the conformal block by
\bea
\label{LioCB}
{\cal F}_{\alpha}^{\lambda}(q)
=
\left(q^{-\frac{1}{24}}\eta(q)\right)^{-1}
H(\lambda,\Delta_\alpha|q)\,,
\eea
where $\eta(q)$ is Dedekind's eta function,
\bea
\eta(q)=q^{\frac{1}{24}}\prod_{n=1}^{\infty}(1-q^n)\,.
\eea
The function $H(\lambda,\Delta_\alpha|q)$ is then determined by the recursion
relation
\bea
\label{RecRel}
H(\lambda,\Delta_\alpha|q)
=
1+
\sum_{m,n=1}^{\infty}
\frac{q^{mn}R_{m,n}}
{\Delta_\alpha-\Delta_{m,n}}\,
H(\lambda,\Delta_{m,n}+mn|q)\,.
\eea
The first terms in the $q$-expansion are
\bea
\label{Hser}
&H(\lambda,\Delta_\alpha|q)
=
1+
\frac{q R_{1,1}}{\Delta_\alpha -\Delta _{1,1}}+
q^2 \left(\frac{R_{1,1}^2}{\Delta_\alpha -\Delta _{1,1}}+\frac{R_{1,2}}{\Delta_\alpha -\Delta _{1,2}}+\frac{R_{2,1}}{\Delta_\alpha -\Delta _{2,1}}\right)
+O(q^3)\,.
\eea
Here $\Delta_{m,n}$ denotes the degenerate conformal dimension
\bea
\Delta_{m,n}
=
\frac{1}{4}\left(Q^2-Q_{m,n}^2\right)\,,
\eea
where
\bea
\label{Qmn}
Q_{m,n}
=
mb+\frac{n}{b}.
\eea
The residues are given by
\bea
\label{R}
R_{m,n}
=
\frac{\prod_{r,s}
	\left(Q_{r,s}-\lambda\right)
	\left(Q_{r,s}-\lambda+b\right)
	\left(Q_{r,s}-\lambda+Q\right)
	\left(Q_{r,s}-\lambda-b+Q\right)}{2 \prod_{k,l}^{\prime} Q_{k,l}}\,.
\eea
The products are taken over the ranges
\bea
\nonumber
& r=-m+1,-m+3,\ldots,m-1\,,
\\
\nonumber
& s=-n+1,-n+3,\ldots,n-1\,,
\\
\nonumber
& k=-m+1,-m+2,\ldots,m-1,m\,,
\\
\nonumber
& l=-n+1,-n+2,\ldots,n-1,n\,.
\eea
Notice that $r$ and $s$ are shifted in steps of two, whereas $k$ and $l$ are
shifted in steps of one. 
The prime in the product in the definition of $R_{m,n}$ indicates that the two pairs $(k,l)=(0,0)$ and $(k,l)=(m,n)$ are omitted.

Before proceeding, let us also recall the result of  \cite{Perlmutter:2015iya}, where Zamolodchikov's recursion relation was used to obtain a closed-form representation for the expansion coefficients of the four-point conformal block. As noted there, the same method applies to the torus one-point conformal block as well. Let us briefly review this construction.
We rewrite (\ref{RecRel}) as
\bea
\label{HinZetaL}
H(\lambda,\Delta_\alpha|q)
=
1+
\sum_{m,n=1}^{\infty}
q^{mn}
\zeta_{m,n}(\lambda,\Delta_\alpha)
H(\lambda,\Delta_{m,n}+mn|q)\,,
\eea
where
\bea
\label{zetaL}
\zeta_{m,n}(\lambda,\Delta_\alpha)=
\frac{R_{m,n}}
{\Delta_\alpha-\Delta_{m,n}}\,.
\eea
The function $H(\lambda,\Delta_\alpha|q)$ admits a power series expansion in $q$,
\bea
\label{HinHN}
H(\lambda,\Delta_\alpha|q)=
1+\sum_{N=1}^{\infty}
H_N(\lambda,\Delta_\alpha)\,q^N\,.
\eea
Following \cite{Perlmutter:2015iya}, iterating the recursion yields a closed-form expression for the coefficients $H_N(\lambda,\Delta_\alpha)$:
\bea
\label{HN}
H_N(\lambda,\Delta_\alpha)=
\sum_{j=1}^N
\bigg[
\prod_{\ell=1}^j \sum_{m_\ell,n_\ell \ge 1}
\zeta_{m_\ell,n_\ell}(\lambda,\Delta^{(\ell)}_{\alpha,{\rm eff}})
\bigg]
\bigg|_{\sum_{\ell=1}^jm_\ell n_\ell=N}\,.
\eea
Here the $j$-th term collects the contributions of $j$ successive applications of the recursion, the sums over $m_\ell,n_\ell$ are restricted by the constraint $\sum_{\ell=1}^j m_\ell n_\ell=N$, and the effective dimensions are defined recursively by
\bea
\label{DefDelEll}
\Delta^{(\ell)}_{\alpha,{\rm eff}}=\Delta_{m_{\ell-1},n_{\ell-1}}+m_{\ell-1}n_{\ell-1}\,,
\qquad \ell>1\,,
\eea
with the initial condition
\bea
\Delta^{(1)}_{\alpha,{\rm eff}}=\Delta_\alpha\,.
\eea
\subsection{Light-limit recursion}
\label{Llight}
In this subsection, we use the  $\Delta$-recursion relation (\ref{RecRel}) to derive the one-point conformal
block in the light asymptotic limit. 
The light limit is a classical limit in which the central charge is sent to infinity by taking $b \to 0$, while keeping the conformal dimensions finite. 
The latter condition can be achieved by taking the charge parameters to scale as
\bea
\label{lightCh}
\alpha = b\,\tilde{\alpha}\,, \qquad
\lambda = b\,\tilde{\lambda}\,.
\eea
It then follows from (\ref{dim}) that
\bea
\Delta_\alpha 
\mathrel{\underset{b\to 0}{=}}
\tilde{\alpha}\,,
\qquad 
\Delta_\lambda 
\mathrel{\underset{b\to 0}{=}}
\tilde{\lambda}\,.
\eea 
By inspecting (\ref{R}), one finds that
\bea
R_{m,n}
\mathrel{\underset{b\to 0}{=}}
\tilde{R}_{m,n}+ O(b^2)\,.
\eea
On the other hand, it is straightforward to see that
\bea
\Delta_\alpha-\Delta_{m,n}=\frac{\left(b^2 (-2 \tilde{\alpha} +m+1)+n+1\right) \left(b^2 (2 \tilde{\alpha} +m-1)+n-1\right)}{4 b^2}\,.
\eea
These estimates show that, in the light limit of (\ref{RecRel}), the only nonzero contributions come from terms with $n=1$. 
For these terms, we have
\bea
\Delta_\alpha-\Delta_{m,1} \mathrel{\underset{b\to 0}{=}}  \tilde{\alpha} +\frac{m-1}{2}
\eea
and
\bea
\tilde{R}_{m,1}(\tilde{\lambda})=-\frac{(1-\tilde{\lambda})_m (\tilde{\lambda})_m}{2 \; m! (m-1)!}\,.
\eea
Therefore, the recursion relation (\ref{RecRel})   in the light limit becomes
\bea
\label{RecRel_light}
\tilde{H}(\tilde{\lambda},\tilde{\alpha}|q)
=
1+
\sum_{m=1}^{\infty}
\frac{q^{m}\tilde{R}_{m,1}(\tilde{\lambda})}
{ \tilde{\alpha} +\frac{m-1}{2}}\,
\tilde{H}\left(\tilde{\lambda},\frac{1+m}{2}\big{|}q\right)\,.
\eea
The first few terms are
\bea
\tilde{H}(\tilde{\lambda},\tilde{\alpha}|q)
&=&
1+q\frac{ \tilde{R}_{1,1}}{\tilde{\alpha} }
+
q^2\left(\frac{\tilde{R}_{1,1}^2}{\tilde{\alpha} } +\frac{2 \tilde{R}_{2,1}}{2 \tilde{\alpha} +1} \right)
\\ \nonumber
&+&
q^3\left(
\frac{\tilde{R}_{1,1}^3}{\tilde{\alpha} }
+
\frac{2 \tilde{R}_{1,1} \tilde{R}_{2,1}}{3 \tilde{\alpha} }
+
\frac{4 \tilde{R}_{1,1} \tilde{R}_{2,1}}{3 (2 \tilde{\alpha} +1)}
+
\frac{2 \tilde{R}_{3,1}}{2 \tilde{\alpha} +2}
\right)
+ O(q^4)\,.
\eea
Thus, we have obtained a recursion relation for the conformal block (\ref{LioCB}) in the light limit:
\bea
{\cal F}_{\alpha}^{\lambda}(q)
\mathrel{\underset{b\to 0}{=}}
\left(q^{-\frac{1}{24}}\eta(q)\right)^{-1}
\tilde{H}(\tilde{\lambda},\tilde{\alpha}|q)\,.
\eea
On the other hand, (\ref{RecRel_light}) can be solved explicitly; the result
is\footnote{The proof that the solution is this hypergeometric function
	follows the same steps as in Appendix~\ref{h}, since the parameters of
	the hypergeometric function in (\ref{tildHhyp})  are related to each other exactly as those of
	(\ref{hsol}).}
\bea
\label{tildHhyp}
\tilde{H}(\tilde{\lambda},\tilde{\alpha}|q)=
(1-q)^{\tilde{\lambda }} \, _2F_1(\tilde{\lambda },2 \tilde{\alpha}+\tilde{\lambda }-1;2 \tilde{\alpha };q)\,.
\eea
Thus, we obtain the known result that the Liouville light one-point conformal block is given by
\bea
{\cal F}_{\alpha}^{\lambda}(q)
\mathrel{\underset{b\to 0}{=}}
\left(q^{-\frac{1}{24}}\eta(q)\right)^{-1}(1-q)^{\tilde{\lambda }}
\, _2F_1(\tilde{\lambda },2 \tilde{\alpha}+\tilde{\lambda }-1;2 \tilde{\alpha };q)\,.
\eea
The closed-form expression (\ref{HN}) for the expansion coefficients also admits a
light limit. In this limit the expansion (\ref{HinHN}) becomes
\bea
\label{HinHNLIGHT}
\tilde{H}(\tilde{\lambda},\tilde{\alpha}|q)=
1+\sum_{N=1}^{\infty}
\tilde{H}_N(\tilde{\lambda},\tilde{\alpha})\,
q^N\,.
\eea
Since the light recursion relation (\ref{RecRel_light}) has the same structure as the
generic one (\ref{RecRel}), with only the terms $n=1$ retained, the coefficients are
given by a  similar expression 
\bea
\label{HNlight}
\tilde{H}_N(\tilde{\lambda},\tilde{\alpha})=
\sum_{j=1}^N
\bigg[ \prod_{\ell=1}^j \sum_{m_\ell \ge 1}
\tilde{\zeta}_{m_\ell,1}(\tilde{\lambda},\tilde{\alpha}^{(\ell)}_{{\rm eff}})
\bigg] \bigg|_{\sum_{\ell=1}^jm_\ell=N}\,,
\eea
where
\bea
\tilde{\zeta}_{m,1}(\tilde{\lambda},\tilde{\alpha})
=
\frac{\tilde{R}_{m,1}(\tilde{\lambda})}
{ \tilde{\alpha} +\frac{m-1}{2}}\,,
\eea
and the effective arguments are
\bea
\label{tildeAlphaEll}
&&\tilde{\alpha}^{(1)}_{{\rm eff}}=\tilde{\alpha}\,,
\\
\nonumber
&&\tilde{\alpha}^{(\ell)}_{{\rm eff}}=
\frac{m_{\ell-1}+1}{2}\,,
\qquad
\ell>1\,.
\eea
In the Liouville case this representation offers
little new information, since the coefficients of (\ref{tildHhyp}) are already known
in a simpler form. The situation is different for $A_2$ Toda theory, where no such
simple explicit solution is available; there we shall write the analog of (\ref{HNlight})
for both the generic and the light block.
\section{$A_2$ Toda Theory}
\label{TodaLight}

\subsection{ Zamolodchikov-like recursion for the $A_2$ Toda torus block}

This subsection reviews $A_2$ Toda conformal field theory \cite{Zamolodchikov:1985wn} and the Zamolodchikov-like $\Delta$-recursion relation \cite{Poghossian:2017atl}, which will later be used to derive the light one-point torus conformal block.
$A_2$ Toda theory is a two-dimensional CFT which, in addition to the spin-two holomorphic
energy-momentum tensor $W^{(2)}(z)\equiv T(z)$, possesses an additional 
spin-three current $W^{(3)}(z)$. 
Its Virasoro central charge is
\bea
c=2+24 Q^2\,,
\quad 
Q=b+b^{-1}\,,
\eea
where $b$ is the dimensionless Toda coupling.

In this paper, we represent roots, weights, and Cartan elements of the Lie algebra $A_2$ as
three-component vectors whose components sum to zero.
The scalar product is taken to be the standard Euclidean one. 
This is equivalent to the more conventional representation of these quantities as diagonal
traceless $3 \times 3$ matrices, with pairing given by the trace of matrix product.
In this representation, the Weyl vector is
\bea
\bm{\rho}=(1,0,-1)\,.
\eea
For later reference, let us give the explicit expressions for the highest weight
$\bm{\omega}_1$ of the first fundamental representation and for its complete set of weights
$\bm{h}_1, \ldots, \bm{h}_3$:
\bea
&&\bm{\omega}_1=\left(\frac{2}{3},-\frac{1}{3},-\frac{1}{3}\right)\,,\nonumber\\
&&(\bm{h}_l)_k=\delta_{l,k}-\frac{1}{3}\,.
\eea
Notice that $\bm{h}_1=\bm{\omega}_1$.

The primary fields $V_{\bm{\alpha}}$ are parametrized by vectors $\bm{\alpha}$ with
vanishing sum of components. 
Here we restrict ourselves to the left-moving holomorphic sector. 
Their conformal weights are given by
\bea
\Delta_{\bm{\alpha}}=\frac{\bm{\alpha}\cdot(2 Q \bm{\rho}-\bm{\alpha})}{2}\,.
\label{dim_gen}
\eea
In what follows, we will consider torus one-point block where the insertion field is a partially degenerate primary field $V_{\lambda \bm{\omega}_1}$,
with conformal dimension 
\bea
\Delta_{\lambda \bm{\omega}_1}=\lambda\left(Q-\frac{\lambda}{3}\right)\,.
\label{dim_ext}
\eea
Sometimes 
it is more convenient to use the  ``momentum''  vector
\bea
\bm{p}=Q \bm{\rho}-\bm{\alpha}\,.
\eea
Besides their conformal dimensions, the fields are in addition  characterized by the eigenvalue
of the zero mode of the $W^{(3)}(z)$ current,
\bea
\label{w}
w=-\frac{i}{27}\sqrt{\frac{48}{22+5c}}\; v\,,
\eea
where $v$ is defined in terms of the momentum vector $\bm{p}$ as
\bea
v=27 p_1 p_2 p_3=(p_{12}-p_{23})(p_{12}+2p_{23})(2 p_{12}+p_{23})\,.
\eea
Another useful combination is
\bea
u=p_{12}^2+p_{23}^2+p_{12}p_{23}\,,
\eea
in terms of which the Virasoro dimension of the primary field $V_{\bm{\alpha}}$ is expressed as
\bea
\Delta_{\bm{\alpha}}=Q^2-\frac{u}{3}\,.
\eea
The one-point conformal block on the torus with modular parameter $\tau$ is defined as
\footnote{In this paper, we deal exclusively with the insertion of a partially degenerate primary
	field whose charge is proportional to the first fundamental weight;
	the treatment of more general cases is substantially more complicated \cite{Fateev:2007ab,Fateev:2008bm,Fateev:2011qa}.}
\bea
\label{1pBlock}
{\cal F}_{\bm{\alpha}}^\lambda (q)=q^{\frac{c}{24}-\Delta_{\bm{\alpha}}} {\rm tr}_{\bm{\alpha}}
\left(q^{L_0-\frac{c}{24}} V_{\lambda \bm{\omega}_1}(1)\right)\,.
\eea
Readers interested in a more detailed explanation of the definition of this conformal block can find it, for example, in \cite{Belavin:2024nnw}.
In \cite{Poghossian:2017atl}, a Zamolodchikov-like $\Delta$-recursion relation was written for this block. 
Let us briefly review it. 
One first factors out a simple prefactor from the conformal block (\ref{1pBlock}) as follows:
\bea
\label{A2block_H}
{\cal F}_{\bm{\alpha}}^\lambda (q)=\left(q^{-\frac{1}{24}}\eta(q)\right)^{-2}H(v^2,u|q)\,.
\eea
The recursion relation then takes the form
\bea
\label{Rectoda}
H(v^2,u|q)=1+\sum_{m,n=1}^\infty \frac{q^{m n} R_{m,n}(u)}{v^2-v_{m,n}^2(u)}
H\left(v^2_{m,-n}(u-3mn),u-3mn|q\right)\,.
\eea
The first terms in the $q$-expansion generated by this recursion are
\bea
H(v^2,u|q)&=&
1+\frac{q R_{1,1}(u)}{v^2-v^2_{1,1}(u)}
+q^2\left(
\frac{R_{1,2}(u)}{v^2-v^2_{1,2}(u)}+\frac{R_{2,1}(u)}{v^2-v^2_{2,1}(u)}
\right.
\\ \nonumber
&+&\left.\frac{R_{1,1}(u-3) R_{1,1}(u)}{\left(v^2_{1,-1}(u-3)-v^2_{1,1}(u-3)\right) \left(v^2-v^2_{1,1}(u)\right)}
\right)+O(q^3)\,,
\eea
where
\bea
\nonumber
R_{m,n}=-18\lambda Q_{m,n}\left(u-Q^2_{m,n}\right)\left(u-3Q^2_{m,n}\right)
\prod_{i=1-m}^{m}
\sideset{}{'}\prod_{j=1-n}^{n}
\frac{Q_{i,j}-\frac{\lambda}{3}}{Q_{i,j}}
\\
\label{Rtoda}
\times	\prod_{i=1}^{m}	\prod_{j=1}^{n}
\frac{		u-Q_{m,n}^2+
	\left(Q_{i,j}-\frac{\lambda}{3}\right)
	\left(Q_{m-i,n-j}+\frac{\lambda}{3}\right)	}
{		u-Q_{m,n}^2+	Q_{i,j}Q_{m-i,n-j}	}\,,
\eea
with
\bea
\label{vnm}
v_{m,n}^2(u)=\left(3Q_{m,n}^2-u\right)^2\left(4u-3Q_{m,n}^2\right)\,.
\eea
In the expression (\ref{Rtoda}), the prime on the product means that the term with $(i,j)=(0,0)$ is omitted.
Finally, $Q_{m,n}$ is given by (\ref{Qmn}). 

As in the Liouville case, the coefficients of the $q$-expansion can be written in
closed form. To this end, we rewrite the recursion relation (\ref{Rectoda}) as
\bea
\label{HNW31}
H(v^2,u|q)
=
1+\sum_{m,n=1}^{\infty} q^{mn}\,
\zeta_{m,n}(v^2,u)\,
H\left(v^2_{m,-n}(u-3mn),u-3mn|q\right)\,,
\eea
with
\bea
\label{zetaW3}
\zeta_{m,n}(v^2,u)=\frac{R_{m,n}(u)}{v^2-v_{m,n}^2(u)}\,.
\eea
The recursion (\ref{HNW31}) generates the power series
\bea
\label{HNW32}
H(v^2,u|q)=1+\sum_{N=1}^{\infty}H_N(v^2,u)\,q^N\,.
\eea
The relations (\ref{HNW31}) and (\ref{zetaW3}) closely resemble their Liouville counterparts
(\ref{HinZetaL}) and (\ref{zetaL}). The only difference is that each application of
the recursion now shifts both arguments of $H$: the parameter $u$ is shifted by
$-3mn$, while $v^2$ is replaced by $v^2_{m,-n}$ evaluated at the shifted value
of $u$. Guided by this structure, and by computing several terms explicitly from
(\ref{HNW31}), we find that
\bea
\label{HNtoda}
H_N(v^2,u)=
\sum_{j=1}^N
\bigg[\prod_{\ell=1}^j
\sum_{m_\ell,n_\ell \ge 1}
\zeta_{m_\ell,n_\ell}\!\left(v^{2\,(\ell)}_{\rm eff},\,u^{(\ell)}_{\rm eff}\right)
\bigg]
\bigg|_{\sum_{\ell=1}^jm_\ell n_\ell=N}\,,
\eea
where the effective arguments are given by
\bea
\label{ueff}
&&u^{(\ell)}_{\rm eff}=u-3\sum_{i=1}^{\ell-1}m_i n_i\,,
\\
\nonumber
&&v^{2\,(1)}_{\rm eff}=v^2\,,
\\
\nonumber
&&v^{2\,(\ell)}_{\rm eff}=v^2_{m_{\ell-1},-n_{\ell-1}}\!\left(u^{(\ell)}_{\rm eff}\right)\,,
\qquad
\ell > 1\,.
\eea
\subsection{The light asymptotic limit}
\label{Tlight}
We now derive the light-limit form of the recursion relation (\ref{Rectoda}).
For $A_2$ Toda theory, this limit is reached by assigning the following $b$-dependence to the charge parameters:
\bea
\alpha_i = b\,\tilde{\alpha}_i\,, \qquad
\lambda = b\,\tilde{\lambda}\,.
\eea
This choice ensures that the conformal dimensions of the fields remain finite as $b\to 0$.
More precisely,
\bea
\Delta_{\bm{\alpha}}
\mathrel{\underset{b\to 0}{=}}
\tilde{\Delta}=
2 \tilde{\alpha}_1+\tilde{\alpha}_2\,.
\eea
From (\ref{w}) we also find
\bea
\label{wL}
w
\mathrel{\underset{b\to 0}{=}}
-i \sqrt{\frac{2}{5}}\tilde{\alpha}_2\,.
\eea
Let us now describe how the recursion relation for the conformal block (\ref{A2block_H}) is obtained in the light asymptotic limit. In this limit, the block takes the form
\bea
\label{Lblock}
{\cal F}_{\bm{\alpha}}^\lambda (q)
\mathrel{\underset{b\to 0}{=}}
\left(q^{-\frac{1}{24}}\eta(q)\right)^{-2}
\tilde{H}(\tilde{\alpha}_2^2,\tilde{\Delta}|q)\,.
\eea
From (\ref{Rtoda}) we observe that
\bea
\label{Rm1ser}
R_{m,1}(u+3k)&=&
\frac{ 3^6 }{b^2}\tilde{R}_{m,1}(\tilde{\Delta}-k)+\ldots\,,
\\
\label{Rm2ser}
R_{m,2}(u+3k)&=&\frac{ 3^6 }{b^4}\tilde{R}_{m,2}+\ldots\,,
\eea
where $\tilde{R}_{m,1}$ and $\tilde{R}_{m,2}$ are defined in (\ref{Rm1}) and (\ref{Rm2}), respectively, and $k$ is a $b$-independent constant.
For $n>2$ one has $R_{m,n}(u)=O\left(b^{-4}\right)$.
On the other hand,
\bea
\label{debehavior}
v^2-v_{m,n}^2(u)=\frac{27 \left(n^2-4\right) \left(n^2-1\right)^2}{b^6}+O\left(b^{-4}\right)\,.
\eea
Thus, in the light limit only the terms with $n=1,2$ contribute to (\ref{Rectoda}).
A closer look at these two cases shows that
\bea
\label{V1}
v^2_{m,\pm 1}(u+3k) &=&
\frac
{3^6}{b^2}\tilde{v}^2_{m,\pm 1}(\tilde{\Delta}-k)+\ldots\,,
\\
\label{V2}
v^2_{m,\pm 2}(u+3k)
&=&
\frac{ 3^6 }{b^4}
\tilde{v}^2_{m,\pm 2}(\tilde{\Delta}-k)
+\ldots\,,
\eea
where $\tilde{v}^2_{m,\pm 1}$ and $\tilde{v}^2_{m,\pm 2}$ are defined in (\ref{vmn}). We also have
\bea
\label{Vser}
v^2&=&\frac{3^6 \tilde{\alpha }^2_2}{b^2}+\ldots\,.
\eea
Using the expansions (\ref{Rm1ser})--(\ref{Vser}), one can derive
the light limit of the recursion relation (\ref{Rectoda})\footnote{One can also show that $H\big(v^2_{m,-n}(u-3mn),u-3mn,\big|,q\big)$
	remains bounded as $b \to 0$.}; more details on the derivation are given in Appendix \ref{app}. The final result reads
\bea
\label{RectodaLight0}
h(x,\tilde{\Delta}|q)=1+
\sum_{m=1}^\infty \frac{q^{2 m } \tilde{R}_{m,2}}{x-\tilde{v}_{m,2}^2(\tilde{\Delta})}
h\left(\tilde{v}^2_{m,-2}\left(\tilde{\Delta}+2 m \right),\tilde{\Delta}+2 m|q\right)\,,
\\
\label{RectodaLight}
\tilde{H}(\tilde{\alpha}_2^2,\tilde{\Delta}|q)
=h(0,\tilde{\Delta}|q)+
\sum_{m=1}^\infty \frac{q^{m } \tilde{R}_{m,1}(\tilde{\Delta})}{\tilde{\alpha}_2^2-\tilde{v}_{m,1}^2(\tilde{\Delta})}
\tilde{H}\left(\tilde{v}^2_{m,-1}\left(\tilde{\Delta}+m \right),\tilde{\Delta}+m|q\right)\,.
\eea
Here
\bea
\label{vmn}
\tilde{v}^2_{m,n}(\tilde{\Delta})=
4 | n|  \left(\frac{1-\frac{\tilde{\Delta}}{2}-\frac{m}{n}}{3} \right)^{\frac{2}{| n| }}
\eea
and
\bea
\label{Rm1}
&&\tilde{R}_{m,1}(\tilde{\Delta})=\left(\frac{2}{3}\right)^2\frac{ \left(\tilde{\Delta}+2m-2\right) \left(1-\frac{\tilde{\lambda}}{3}\right)_m \left(\frac{\tilde{\lambda }}{3}\right)_m}{  (m-1)! m!}\,,
\\
\label{Rm2}
&&\tilde{R}_{m,2}=-\frac{4 \left(1-\frac{\tilde{\lambda }}{3}\right)_m \left(\frac{\tilde{\lambda }}{3}\right)_m}{3(m-1)! m!}\,.
\eea
Notice that the relation (\ref{RectodaLight0}) determines the function $h(x,\tilde{\Delta}|q)$ uniquely as a power series in $q$.
This recursion can be solved explicitly, with the result
\bea
\label{hsol}
h(x,\tilde{\Delta}|q)
=\left(1-q^2\right)^{\frac{\tilde{\lambda}}{3}}
\, _2F_1\left(\frac{\tilde{\lambda }}{3},\frac{\tilde{\lambda }}{3}+\frac{3 x}{4}+\tilde{\Delta} -2;\frac{3 x}{4}+\tilde{\Delta} -1;q^2\right)
\,;
\eea
a proof that (\ref{hsol}) satisfies the recursion (\ref{RectodaLight0}) is given in Appendix \ref{h}.
Inserting this result into (\ref{RectodaLight}), we arrive at the final form of the recursion relation given in Subsection~\ref{WrecLight}.

Let us now write down the light-limit analogues of (\ref{HNW32}) and (\ref{HNtoda}). Comparing (\ref{RectodaLight}) with (\ref{Rectoda}), we deduce that
\bea
\label{Ht2}
\tilde{H}(\tilde{\alpha}_2^2,\tilde{\Delta}|q)=1+\sum_{N=1}^{\infty}\tilde{H}_N(\tilde{\alpha}_2^2,\tilde{\Delta})\,q^{N}\,,
\eea
with
\bea
\label{HNlightW}
\tilde{H}_N(\tilde{\alpha}_2^2,\tilde{\Delta})=
\sum_{j=1}^{N}
\bigg[\prod_{\ell=1}^{j}\;
\sum_{m_\ell\ge 1}\;\sum_{n_\ell=1}^{2}
\tilde{\zeta}_{m_\ell,n_\ell}\!\left(\tilde{v}^{2\,(\ell)}_{\rm eff},\,\tilde{\Delta}^{(\ell)}_{\rm eff}\right)
\bigg]
\bigg|_{\substack{\sum_{\ell=1}^{j}m_\ell n_\ell=N\\[2pt] n_1\le n_2\le\cdots\le n_j}}\,,
\eea
and
\bea
\tilde{\zeta}_{m,n}(x,\tilde{\Delta})=\frac{\tilde{R}_{m,n}(\tilde{\Delta})}{x-\tilde{v}^2_{m,n}(\tilde{\Delta})}\,,
\qquad n=1,2\,,
\eea
where the effective arguments are given by
\bea
&&\tilde{\Delta}^{(\ell)}_{\rm eff}=\tilde{\Delta}+\sum_{i=1}^{\ell-1}m_i n_i\,,
\nonumber\\
&&\tilde{v}^{2\,(1)}_{\rm eff}=\delta_{n_1,1}\,\tilde{\alpha}_2^2\,,
\nonumber\\
&&\tilde{v}^{2\,(\ell)}_{\rm eff}=\delta_{n_\ell,n_{\ell-1}}\,
\tilde{v}^2_{m_{\ell-1},-n_{\ell-1}}\!\left(\tilde{\Delta}^{(\ell)}_{\rm eff}\right)\,,
\qquad \ell>1\,.
\eea
The ordering condition $n_1\le n_2\le\cdots\le n_j$ reflects the structure of the coupled recursion:
a chain of steps generated by (\ref{RectodaLight}) (with $n_\ell=1$) may be followed by steps generated by (\ref{RectodaLight0})
(with $n_\ell=2$), but not vice versa. The Kronecker symbols account for the fact that the function $h$ enters
(\ref{RectodaLight}) with vanishing first argument.
\subsection{The light $\Delta$-recursion}
\label{WrecLight}
Let us summarize the main result of this paper. The Zamolodchikov-like recursion for the  $A_2$ Toda light torus block is
\bea
\nonumber
\tilde{H}(\tilde{\alpha}_2^2,\tilde{\Delta}|q)
&=&
\left(1-q^2\right)^{\frac{\tilde{\lambda }}{3}}
\, _2F_1\left(\frac{\tilde{\lambda }}{3},\frac{\tilde{\lambda }}{3}+\tilde{\Delta}-2;\tilde{\Delta}-1;q^2\right)
\\ \label{RectodaLightMAIN}
&+&
\sum_{m=1}^\infty \frac{q^{m } \tilde{R}_{m,1}(\tilde{\Delta})}{\tilde{\alpha}_2^2-\tilde{v}_{m,1}^2(\tilde{\Delta})}
\tilde{H}\left(\tilde{v}^2_{m,-1}\left(\tilde{\Delta}+m \right),\tilde{\Delta}+m|q\right)\,,
\eea
where
\bea
\tilde{v}^2_{m,\pm 1}(\tilde{\Delta})=\left(\frac{2}{3}\right)^2   \left(\frac{\tilde{\Delta}}{2} \pm m-1 \right)^{2}\,,
\\
\tilde{R}_{m,1}(\tilde{\Delta})=\left(\frac{2}{3}\right)^2\frac{ \left(\tilde{\Delta}+2m-2\right) \left(1-\frac{\tilde{\lambda}}{3}\right)_m \left(\frac{\tilde{\lambda }}{3}\right)_m}{  (m-1)! m!}\,.
\eea
We have checked that this recursion relation, inserted into (\ref{Lblock}), is in full agreement with the result of \cite{Poghosyan:2026yow} up to thirtieth order in
 $q$. Since the expression of \cite{Poghosyan:2026yow} was in turn compared with that of \cite{Belavin:2024nnw}, this comparison carries over to our result as well.

\section*{Acknowledgments}
The research of A.P. was supported by the Armenian SCS grant 21AG-1C060. H.P.'s work was supported by the Armenian SCS grants 21AG-1C062
and 24WS-1C031.

\vspace*{3pt}
\appendix
\section{More details on the light limit of the $A_2$ one-point block}
\label{app}
In this appendix we explain the origin of the structure of the light-limit recursion (\ref{RectodaLight0}), (\ref{RectodaLight}), in particular why the $n=1$ and $n=2$ terms of (\ref{Rectoda}) partially decouple in the limit $b\to 0$. Throughout, we use the ordering convention introduced after (\ref{HNlightW}): the outermost factor of an iterated term is the first step, $\ell=1$.

Consider two adjacent factors in the iterated recursion with $n_{\ell-1}=2$ and $n_\ell=1$. Then it follows from (\ref{Rm1ser}), (\ref{V1}) and (\ref{V2}) that
\bea
\label{vanish}
\frac{ R_{m_\ell,1}(u-6m_{\ell-1})}{v_{m_{\ell-1},-2}^2(u-6m_{\ell-1})-v_{m_\ell,1}^2(u-6m_{\ell-1})}
\mathrel{\underset{b\to 0}{=}}
0\,.
\eea
This fact, together with the structure of the recursion (\ref{Rectoda}), shows that in the light limit an $n=1$ factor can never be preceded by an $n=2$ factor, which is precisely the ordering condition $n_1\le n_2\le\cdots\le n_j$ in (\ref{HNlightW}).

This observation has two consequences:
\begin{itemize}
	\item The $n=2$ factors always occupy the last positions of a chain and are never followed by $n=1$ factors. They therefore form a closed subsystem, governed by the separate recursion relation (\ref{RectodaLight0}) for the function $h$.
	\item The $n=1$ factors occupy the first positions, and the transition from $n=1$ to $n=2$ occurs at most once along a chain. This is why the function $h$ appears in (\ref{RectodaLight}).
\end{itemize}

Let us also explain the arguments of the functions appearing in (\ref{RectodaLight}). From (\ref{Vser}) and (\ref{V2}) one immediately sees
that $v^2$ is subleading with respect $v_{m,2}^2(u) $ and
\bea
\label{vvm2}
v^2-v_{m,2}^2(u)&=&-\frac{3^6}{b^4}\tilde{v}_{m,2}^2+\ldots\,,
\eea
Hence, at the transition from the $n=1$ chain to the $n=2$ chain, the first argument of $H$ drops out at leading order, and the function $h$ enters (\ref{RectodaLight}) with vanishing first argument, $x=0$. 

Similarly, from (\ref{Vser}) and (\ref{V1}) we get
\bea
\label{vvm1}
v^2-v_{m,1}^2(u)&=&\frac{3^6}{b^2}\left(\tilde{\alpha}_2^2-\tilde{v}_{m,1}^2\right)+\ldots\,,
\eea
which explains the form of the denominator in the second term on the right-hand side of (\ref{RectodaLight}).
\section{Solving the recursion for $h(x,\tilde{\Delta}|q)$}
\label{h}
To prove that (\ref{hsol}) satisfies the recursion (\ref{RectodaLight0}), it suffices to show that, as functions of $\tilde{\Delta}$, both expressions have the same large-$\tilde{\Delta}$ behavior and the same residues. It is known that
\bea
\, _2F_1\left(\frac{\tilde{\lambda }}{3},\frac{\tilde{\lambda }}{3}+\frac{3 x}{4}+\tilde{\Delta} -2;\frac{3 x}{4}+\tilde{\Delta} -1;q^2\right)
\mathrel{\underset{\tilde{\Delta}\to \infty}{=}}
\left(1-q^2\right)^{-\frac{\tilde{\lambda}}{3}}\,.
\eea
Hence, from (\ref{hsol}) we see that
\bea
h(x,\tilde{\Delta}|q)
\mathrel{\underset{\tilde{\Delta}\to \infty}{=}}
1\,.
\eea
On the other hand, taking into account (\ref{vmn}) and (\ref{Rm2}), one finds the same large-$\tilde{\Delta}$ behavior from (\ref{RectodaLight0}).

Let us now compare the residues. From (\ref{hsol}) we get
\bea
\label{Resh}
\mathrel{\underset{\tilde{\Delta}=-\frac{3 x}{4}-n+1}{{\rm Res}}}h(x,\tilde{\Delta}|q)
&=&
-q^{2(n+1)} \frac{\left(\frac{\tilde{\lambda}}{3}\right)_{n+1}\left(1-\frac{\tilde{\lambda}}{3}\right)_{n+1}}{ n! (n+1)!}
\\ \nonumber
& \times & \left(1-q^2\right)^{\frac{\tilde{\lambda}}{3}}
{}_2F_1\!\left(\frac{\tilde{\lambda}}{3},\, \frac{\tilde{\lambda}}{3}+n+1;\, n+2;\, q^2\right)\,.
\eea
On the other hand, a direct computation of the residue at $\tilde{\Delta}=-\frac{3 x}{4}-n+1$ from (\ref{RectodaLight0}) gives
\bea
&\mathrel{\underset{\tilde{\Delta}=-\frac{3 x}{4}-n+1}{{\rm Res}}}h(x,\tilde{\Delta}|q)
=
- q^{2 (n+1) } \frac{ \left(1-\frac{\tilde{\lambda }}{3}\right)_{n+1} \left(\frac{\tilde{\lambda }}{3}\right)_{n+1}}{ \; n! (n+1)!}\qquad \qquad\\
\nonumber
&\qquad \qquad \times
h\left(\tilde{v}^2_{n+1,-2}\left(-\frac{3 x}{4}+n+3 \right),-\frac{3 x}{4}+n+3|q\right)\,.
\eea
Substituting (\ref{hsol}) for the function $h$ on the right-hand side, this coincides with (\ref{Resh}). 
Combining this result with the fact that, order by order in $q$, both sides are rational in $\tilde\Delta$, 
we conclude that their difference is entire and bounded and hence vanishes. This completes the proof.

	\bibliographystyle{JHEP}
	\providecommand{\href}[2]{#2}\begingroup\raggedright\endgroup
	
\end{document}